%% file: main.tex
\documentclass{easychair}

\usepackage{pstricks}
\usepackage{pst-node}

\newcommand\lmcshevea[2]{#1}

\input{macros}

\newtheorem{defi}{Definition}[section]
\newtheorem{prop}{Proposition}[section]
\newtheorem{theorem}{Theorem}[section]

\newcommand\lmcscorrection[2]{#2}

\title{
    Emptiness of Stack Automata is NEXPTIME-complete: A Correction
    \thanks{We thank an anonymous reviewer for pointing out the error in the original PSPACE algorithm.}
}

\author{
    Christopher Broadbent\inst{1}
    \and
    Arnaud Carayol\inst{2}
    \and
    Matthew Hague\inst{3}\thanks{Supported by EPSRC [EP/K009907/1].}
    \and
    Olivier Serre\inst{4}
}

\institute{
    Institut f\"ur Informatik (I7), Technische Universit\"at M\"unchen
    \and
    Laboratoire d'informatique de l'Institut Gaspard Monge, Universit\'e Paris-Est, and CNRS
    \and
    Royal Holloway, University of London
    \and
    IRIF, Universit\'e Paris Diderot - Paris 7, and CNRS
}

\authorrunning{C. Broadbent, A. Carayol, M. Hague, and O. Serre}

\titlerunning{Emptiness of Stack Automata is NEXPTIME-complete}

\begin{document}

\maketitle

\begin{abstract}
    \input{abstract}
\end{abstract}

\input{introduction}

\input{preliminaries}

\input{emptiness}

\input{conclusion}

\bibliographystyle{plain}
\bibliography{references}

\end{document}

%% file: macros.tex
\usepackage{amsfonts}
\usepackage{amsmath}
\usepackage{amssymb}
\usepackage{xspace}
\usepackage{stmaryrd}
\usepackage{pifont}

\newcommand\neworrenewcommand[1]{%
    \let#1\relax
    \newcommand#1%
}

\providecommand\lmcshevea[2]{#1}

\lmcshevea{
    \providecommand\qed{\hfill\square}
}{
    \providecommand\qed{\square}
}

\newcommand\etal{~\textit{et al.}\xspace}

\newcommand\setcomp[2]{\left\{{#1}\ \left|\ {#2}\right.\right\}}
\newcommand\set[1]{\left\{{#1}\right\}}
\newcommand\tuple[1]{\brac{#1}}
\newcommand\brac[1]{\left({#1}\right)}
\newcommand\apply[2]{{#1}\mathord{\brac{#2}}}
\newcommand\idxi{i}
\newcommand\idxj{j}

\newcommand\numof{\ell}
\lmcshevea{
    \neworrenewcommand\ordinal{\alpha}
}{
    \renewcommand\ordinal{\alpha}
}

\newcommand\naturals{\mathbb{N}}

\newcommand\stacks[1]{{Stacks_{#1}}}

\newcommand\cpds{\mathcal{C}}
\newcommand\controls{\mathcal{P}}
\newcommand\controlset{P}
\newcommand\cpdsrules{\mathcal{R}}
\newcommand\cpdsord{n}
\newcommand\opord{k}
\newcommand\cops[1]{\mathcal{O}_{#1}}

\newcommand\cha{a}
\newcommand\chb{b}
\newcommand\chc{c}
\newcommand\chd{d}
\newcommand\che{e}
\newcommand\cpush[2]{push^{#2}_{#1}}
\newcommand\push[1]{push_{#1}}

\newcommand\pop[1]{pop_{#1}}
\newcommand\collapse[1]{collapse_{#1}}
\newcommand\rew[1]{rew_{#1}}

\newcommand\cpdsrule[4]{\tuple{{#1},{#2},{#3},{#4}}}

\newcommand\control{p}
\newcommand\genop{o}

\newcommand\config[2]{\langle {#1}, {#2} \rangle}

\newcommand\configset{C}
\newcommand\stackw{w}
\newcommand\stacku{u}
\newcommand\stackv{v}

\newcommand\ctop[1]{top_{#1}}
\newcommand\cbottom[2]{bot^{#2}_{#1}}
\newcommand\cpdstran{\longrightarrow}

\newcommand\alphabet\Sigma

\newcommand\ccompose[3]{{#1} :_{#2} {#3}}

\newcommand\annot[2]{{#1}^{#2}}

\newcommand\cpdsalttran[2]{{#1} \rightarrow {#2}}

\newcommand\oalphabet\Gamma
\newcommand\ocha\gamma

\newcommand\substacks[1]{\apply{\mathrm{Subs}}{#1}}

\newcommand\sastates{\mathbb{Q}}
\newcommand\sastateset{Q}
\newcommand\sadelta{\Delta}
\newcommand\safinals{\mathcal{F}}
\newcommand\sastate{q}

\newcommand\saauta{A}
\newcommand\saautb{B}
\newcommand\sopen[1]{[_{#1}}
\newcommand\sclose[1]{]_{#1}}
\newcommand\sbrac[2]{[{#1}]_{#2}}
\newcommand\ssbrac[2]{\left[{#1}\right]_{#2}}
\newcommand\lang{\mathcal{L}}

\newcommand\slang[2]{\apply{\lang_{#1}}{#2}}
\newcommand\satran[1]{\xrightarrow{#1}}
\newcommand\satrancol[2]{\xrightarrow[{#2}]{#1}}

\newcommand\branch{{col}}

\newcommand\tiles{\Theta}
\newcommand\hrel{H}
\newcommand\vrel{V}
\newcommand\tile{t}
\newcommand\inittile{t_I}
\newcommand\fintile{t_F}
\newcommand\tileheight{h}
\newcommand\dimension{N}

\newcommand\spacer{\#}
\newcommand\binof[1]{\langle #1 \rangle}
\newcommand\bone{\hat{1}}
\newcommand\bzero{\hat{0}}

%% file: abstract.tex
A saturation algorithm for collapsible pushdown systems was published in ICALP
2012.
This work introduced a class of \emph{stack automata} used to recognised
regular sets of collapsible pushdown configurations.
It was shown that these automata form an effective boolean algebra, have a
linear time membership problem, and are equivalent to an alternative automata
representation appearing in LICS 2010.
It was also claimed that the emptiness problem for stack automata is
PSPACE-complete.
Unfortunately, this claim is not true.
We show that the problem is in fact NEXPTIME-complete when the stacks being accepted are collapsible pushdown stacks, rather than the annotated stacks used in ICALP 2012.

%% file: introduction.tex
\section{Outline}

We begin with the preliminaries in Section~\ref{sec:preliminaries}.
The complexity of emptiness checking is shown in Section~\ref{sec:emptiness}.

%% file: preliminaries.tex
\section{Preliminaries}
\label{sec:preliminaries}

\lmcscorrection{
    \subsection{Collapsible Pushdown Systems}

    We give the definition of higher-order collapsible stacks and their operations,
    before giving the definition of collapsible pushdown systems.

    \subsubsection{Higher-Order Collapsible Stacks}
}{
    We give the definition of higher-order collapsible stacks before describing
    stack automata.

    \subsection{Higher-Order Collapsible Stacks}
}

Higher-order collapsible stacks are a nested ``stack-of-stacks'' structure
over a stack alphabet $\alphabet$.  Each stack character
contains a pointer --- called a ``link'' --- to a position lower down in the
stack.  The stack operations, defined below, create copies of
sub-stacks.  The link is intuitively a pointer to the context in
which the stack character was first created.
These links will be defined as tuples, the meaning of which is expanded
upon after the following definition.
Let the natural numbers $\naturals$ be
$\set{0, 1, 2, \ldots}$.

\begin{defi}[Order-$\cpdsord$ Collapsible Stacks]
    An \emph{order-$\opord$ link} is a tuple
    $\tuple{\opord, \idxi}$
    where $\opord \geq 1$ and $\idxi$ are natural numbers.
    If
    $\opord \in \set{1, \ldots, \cpdsord}$
    we say the link is \emph{up-to} order-$\cpdsord$.
    Given a finite set of stack characters $\alphabet$, an \emph{order-$0$
    stack} with an up-to order-$\cpdsord$ link is
    $\annot{\cha}{\tuple{\opord, \idxi}}$
    where $\cha \in \alphabet$ and $\tuple{\opord, \idxi}$ is an up-to order-$\cpdsord$ link.
    An \emph{order-$\opord$ stack} with up-to order-$\cpdsord$ links is a sequence
    $\stackw = \sbrac{\stackw_1 \ldots \stackw_\numof}{\opord}$
    such that each $\stackw_\idxj$ is an order-$(\opord-1)$ stack with up-to order-$\cpdsord$ links.
    Moreover, for each $\stackw_\idxj$ and each order-$\opord$ link
    $\tuple{\opord, \idxi}$
    appearing on a character in $\stackw_\idxj$, we have
    $\idxi < \idxj$.
    Let $\stacks{\cpdsord}$ denote the set of order-$\cpdsord$ stacks with up-to order-$\cpdsord$ links.
\end{defi}

In the sequel we will refer to order-$\cpdsord$ stacks with up-to order-$\cpdsord$ links simply as order-$\cpdsord$ stacks.
We will use order-$\opord$ stack to mean an order-$\opord$ stack with up-to order-$\cpdsord$ links, where $\cpdsord$ is clear from the context.
We define the interpretation of the collapse links below.
First, we give an example order-$3$ stack:
$\sbrac{
    \sbrac{\sbrac{\cha^{\tuple{3,1}} \chb^{\tuple{1,0}}}{1}}{2}
    \sbrac{
        \sbrac{\chc^{\tuple{2,1}}}{1}
        \sbrac{\chd^{\tuple{1,1}} \che^{\tuple{1,0}}}{1}
    }{2}
}{3}$.
Intuitively, the collapse links point to a position lower down in the stack.
Hence, we can represent collapse links informally with arrows.
Our example stack could be written
{%
    \vspace{2ex}
    \[
        \sbrac{\quad
            \sbrac{\quad\sbrac{\quad\rnode{a}{\cha}\quad\rnode{b}{\chb}\ \ \pnode{d2}\ \ }{1}\quad}{2}
            \ \ \pnode{d1}\ \ %
            \sbrac{\quad%
                \sbrac{\quad\rnode{c}{\chc}\quad}{1}%
                \ \ \pnode{d3}\ \ %
                \sbrac{\quad\rnode{d}{\chd}\ \ \pnode{d4}\ \ \rnode{e}{\che}\ \ \pnode{d5}\ \ }{1}\quad%
            }{2}\quad%
        }{3} \ .
    \]
    \psset{arcangleA=100,arcangleB=80,nodesep=1pt}%
    \ncarc[ncurv=.7]{->}{a}{d1}%
    \ncarc[ncurv=1.5]{->}{b}{d2}%
    \ncarc[ncurv=1.2]{->}{c}{d3}%
    \ncarc[ncurv=1.5]{->}{d}{d4}%
    \ncarc[ncurv=1.5]{->}{e}{d5}%
}%
The collapse operation, defined below, will remove all parts of the stack above the destination of the topmost collapse link.
Collapse on the stack above gives
$\sbrac{
    \sbrac{
        \sbrac{\chc^{\tuple{2,1}}}{1}
        \sbrac{\chd^{\tuple{1,1}} \che^{\tuple{1,0}}}{1}
    }{2}
}{3}$.
Note, we will often omit the collapse link annotations for readability.

Given an order-$\cpdsord$ stack
$\sbrac{\stackw_1\ldots\stackw_\numof}{\cpdsord}$, we define
\[
    \begin{array}{rcll}
        \apply{\ctop{\cpdsord}}{\sbrac{\stackw_1 \ldots
        \stackw_\numof}{\cpdsord}} &=& \stackw_1 & \text{when $\numof > 0$} \\ 
        \apply{\ctop{\opord}}{\sbrac{\stackw_1 \ldots \stackw_\numof}{\cpdsord}}
        &=& \apply{\ctop{\opord}}{\stackw_1} & \text{when $\opord < \cpdsord$ 
        and $\numof > 0$}
    \end{array}
\]
noting that $\apply{\ctop{\opord}}{\stackw}$ is undefined if
$\apply{\ctop{\opord'}}{\stackw}$ is empty for any $\opord' > \opord$.
For technical reasons, we also define
$\apply{\ctop{\cpdsord+1}}{\stackw} = \sbrac{\stackw}{\cpdsord+1}$
when $\stackw$ is an order-$\cpdsord$ stack.
We remove the top portion of a $\ctop{\opord}$ stack using, where $\idxi > 0$,
\[
    \begin{array}{rcll}%
        \apply{\cbottom{\cpdsord}{\idxi}}{\sbrac{\stackw_1 \ldots %
        \stackw_\numof}{\cpdsord}} &=& \sbrac{\stackw_{\numof - \idxi + 1} %
        \ldots \stackw_\numof}{\cpdsord} & \text{when $\idxi \leq \numof$} \\ %
        \apply{\cbottom{\opord}{\idxi}}{\sbrac{\stackw_1 \ldots %
        \stackw_\numof}{\cpdsord}} &=& %
        \sbrac{\apply{\cbottom{\opord}{\idxi}}{\stackw_1} \stackw_2 \ldots %
        \stackw_\numof}{\cpdsord} & \text{when $\opord < \cpdsord$ and $\numof > %
        0$} \ .%
    \end{array}%
\]

For $\apply{\ctop{1}}{\stackw} = \cha$
where $\cha$ has the link $\tuple{\opord, \idxi}$, the destination of the link
is $\apply{\cbottom{\opord}{\idxi}}{\stackw}$.

When $\stacku$ is a $(\opord-1)$-stack and $\stackv=\sbrac{\stackv_1\ldots\stackv_\numof}{\cpdsord}$ is an $\cpdsord$-stack with $\opord\leq \cpdsord$, we define $\ccompose{\stacku}{\opord}{\stackv}$ as the stack obtained by adding $\stacku$ on top of the topmost $\opord$-stack of $\stackv$. Formally, we let
\[
    \begin{array}{rcll}
        \ccompose{\stacku}{\opord}{\stackv} &=& \sbrac{\stacku\stackv_1\ldots\stackv_\numof}{\cpdsord} & \text{when $\opord= \cpdsord$} \\ 
        \ccompose{\stacku}{\opord}{\stackv} &=& \sbrac{(\ccompose{\stacku}{\opord}{\stackv_1})\stackv_2\ldots\stackv_\numof}{\cpdsord} & \text{when $\opord< \cpdsord$} \\ 
    \end{array}
\]

\lmcscorrection{
    \subsubsection{Operations on Order-$\cpdsord$ Collapsible Stacks}

    The following operations may be performed on an order-$\cpdsord$ collapsible
    stack.
    \[
        \begin{array}{rcl}%
            \cops{\cpdsord} &=& \set{\pop{1},\ldots,\pop{\cpdsord}} \cup %
                                \set{\push{2},\ldots,\push{\cpdsord}}\ \cup \\ %
                            & & \set{\collapse{2},\ldots,\collapse{\cpdsord}} \cup %
                                \setcomp{\cpush{\cha}{2},\ldots,\cpush{\cha}{\cpdsord},\rew{\cha}}{\cha \in \alphabet}%
        \end{array}%
    \]
    We say $\genop \in \cops{\cpdsord}$ is of order-$\opord$ when $\opord$ is
    minimal such that $\genop \in \cops{\opord}$.  E.g., $\push{\opord}$ is
    of order $\opord$.

    The $\collapse{\opord}$ operation is non-standard in the sense of
    Hague\etal~\cite{HMOS08,HMOS17} and has the semantics of a normal collapse, with the
    additional constraint that the top character has an order-$\opord$ link.  The
    standard version of collapse can be simulated with a non-deterministic choice on
    the order of the stack link.  In the other direction, we can store in the stack
    alphabet the order of the collapse link attached to each character on the stack.
    Note, we do not allow order-$1$ links to be created or used.
    In effect, these links are ``null''.

    We define each stack operation in turn for an order-$\cpdsord$ stack $\stackw$.
    Collapse links are created by the $\cpush{\cha}{\opord}$ operations, which add a
    character to the top of a given stack $\stackw$ with a link pointing to
    $\apply{\pop{\opord}}{\stackw}$.
    \begin{enumerate}
        \item We set $\apply{\pop{\opord}}{\stackw} = \stackv$ when $\stackw$
              decomposes into $\ccompose{\stacku}{\opord}{\stackv}$.

        \item We set $\apply{\push{\opord}}{\stackw} =
              \ccompose{\stacku}{\opord}{\ccompose{\stacku}{\opord}{\stackv}}$ when
              $\stackw = \ccompose{\stacku}{\opord}{\stackv}$.

        \item We set $\apply{\collapse{\opord}}{\stackw} =
              \apply{\cbottom{\opord}{\idxi}}{\stackw}$ where
              $\apply{\ctop{1}}{\stackw} = \cha^{\tuple{\opord,\idxi}}$ for some
              $\idxi$.

        \item We set $\apply{\cpush{\chb}{\opord}}{\stackw} =
              \ccompose{\chb^{\tuple{\opord,\numof-1}}}{1}{\stackw}$ where
              $\apply{\ctop{\opord+1}}{\stackw} =
              \sbrac{\stackw_1\ldots\stackw_\numof}{\opord+1}$.

        \item We set $\apply{\rew{\chb}}{\stackw} =
              \ccompose{\chb^{\tuple{\opord,\idxi}}}{1}{\stackv}$ where $\stackw =
              \ccompose{\cha^{\tuple{\opord,\idxi}}}{1}{\stackv}$.

    \end{enumerate}
    Note that, for a $\push{\opord}$ operation, links outside of $\stacku =
    \apply{\ctop{\opord}}{\stackw}$ point to the same destination in both copies of
    $\stacku$, while links pointing within $\stacku$ point within the respective
    copies of $\stacku$.  For full introduction, we refer the reader to
    Hague\etal~\cite{HMOS08,HMOS17}.  In Section~\ref{sec:examples} we give several example
    stacks and show how the stack operations affect them.

    \subsubsection{Collapsible Pushdown Systems}

    We define alternating collapsible pushdown systems.

    \begin{defi}[Collapsible Pushdown Systems]
        An alternating order-$\cpdsord$ \emph{collapsible pushdown system (collapsible PDS)} is
        a tuple $\cpds = \tuple{\controls, \alphabet, \cpdsrules}$ where $\controls$
        is a finite set of control states, $\alphabet$ is a finite stack alphabet,
        and $\cpdsrules \subseteq \brac{\controls \times \alphabet \times
        \cops{\cpdsord} \times \controls} \cup \brac{\controls \times 2^\controls}$
        is a set of rules.
    \end{defi}

    We write \emph{configurations} of a collapsible PDS as a pair $\config{\control}{\stackw}$
    where $\control \in \controls$ and $\stackw \in \stacks{\cpdsord}$.  We write
    $\config{\control}{\stackw} \cpdstran \config{\control'}{\stackw'}$ to denote a
    transition from a rule $\cpdsrule{\control}{\cha}{\genop}{\control'}$ with
    $\apply{\ctop{1}}{\stackw} = \cha$ and $\stackw' = \apply{\genop}{\stackw}$.
    Furthermore, we have a transition $\config{\control}{\stackw} \cpdstran
    \setcomp{\config{\control'}{\stackw}}{\control' \in \controlset}$ whenever we
    have a rule $\cpdsalttran{\control}{\controlset}$.  A non-alternating collapsible PDS has
    no rules of this second form.  We write $\configset$ to denote a set of
    configurations.
}{
}

\subsection{Regularity of Collapsible Stacks}
\label{sec:stackaut}

\lmcscorrection{
    We will present an algorithm that operates on sets of configurations.
}{
    We are interested in regular representations of sets of collapsible
    pushdown stacks.
}
For this we use order-$\cpdsord$ stack automata, thus defining a notion of
regular sets of stacks.  These have a nested structure based on a similar
automata model by Bouajjani and Meyer~\cite{BM04}.  The handling of collapse
links is similar to automata introduced by Broadbent\etal~\cite{BCOS10}, except
we read stacks top-down rather than bottom-up.

\begin{defi}[Order-$\cpdsord$ Stack Automata]
    An \emph{order-$\cpdsord$ stack automaton}
    \[
        \saauta = \tuple{
                      \sastates_\cpdsord,\ldots,\sastates_1,
                      \alphabet,
                      \sadelta_\cpdsord,\ldots,\sadelta_1,
                      \safinals_\cpdsord,\ldots,\safinals_1
                  }
    \]
    is a tuple where
        $\alphabet$ is a finite stack alphabet,
        $\sastates_\cpdsord, \ldots, \sastates_1$ are finite disjoint statessets,
    and
    \begin{enumerate}
        \item
            for all $\opord \in \set{2,\ldots,\cpdsord}$, we have that
            $\sadelta_\opord \subseteq
                \sastates_\opord \times \sastates_{\opord-1} \times 2^{\sastates_\opord}$
            is a transition relation, and
            $\safinals_\opord \subseteq \sastates_\opord$ is a set of accepting states, and

         \item
            $\sadelta_1 \subseteq
                \bigcup\limits_{2 \leq \opord \leq \cpdsord}\brac{
                    \sastates_1 \times
                    \alphabet \times
                    2^{\sastates_\opord} \times
                    2^{\sastates_1}
                }$
            is a transition relation, and
            $\safinals_1 \subseteq \sastates_1$
            a set of accepting states.
    \end{enumerate}
\end{defi}

Stack automata are alternating automata that read the stack in a nested fashion.
Order-$\opord$ stacks are recognised from states in $\sastates_\opord$.  A
transition $\tuple{\sastate, \sastate', \sastateset} \in \sadelta_\opord$ from
$\sastate$ to $\sastateset$ for some $\opord > 1$ can be fired when the
$\ctop{\opord-1}$ stack is accepted from $\sastate' \in \sastates_{(\opord-1)}$.
The remainder of the stack must be accepted from all states in $\sastateset$.
At order-$1$, a transition $\tuple{\sastate, \cha, \sastateset_\branch,
\sastateset}$ is a standard alternating $\cha$-transition with the additional
requirement that the stack pointed to by the collapse link of $\cha$ is accepted
from all states in $\sastateset_\branch$.  A stack is accepted if a subset of
$\safinals_\opord$ is reached at the end of each order-$\opord$ stack.  In
Section~\ref{sec:formal-sa-run}, we formally define the runs of a stack automaton.  We write
$\stackw \in \slang{\sastate}{\saauta}$ whenever $\stackw$ is accepted from a
state $\sastate$. For ease of presentation, we write $\sastate \satran{\sastate'}\sastateset\in \sadelta_\opord$ instead of $\tuple{\sastate, \sastate', \sastateset} \in \sadelta_\opord$ and $\sastate \satrancol{\cha}{\sastateset_\branch} \sastateset
\in \sadelta_1$ instead of $\tuple{\sastate, \cha, \sastateset_\branch,
\sastateset}\in \sadelta_1$.

A (partial) run is informally pictured below, reading an order-$3$ stack using
$\sastate_3 \satran{\sastate_2} \sastateset_3 \in \sadelta_3, \sastate_2
\satran{\sastate_1} \sastateset_2 \in \sadelta_2$ and $\sastate_1
\satrancol{\cha}{\sastateset_\branch} \sastateset_1 \in \sadelta_1$.
Note, the transition
$\sastate_3 \satran{\sastate_2} \sastateset_3$
reads the topmost order-$2$ stack, with the remainder of the stack being read from $\sastateset_3$.
The node labelled $\sastateset_\branch$ begins a run on the stack pointed to by the
collapse link of $\cha$.  Note that the label of this node may contain other
elements apart from $\sastateset_\branch$.  These additional elements come from
the part of the run coming from the previous node (and other collapse links).
\begin{center}
    \vspace{3ex}
    \begin{psmatrix}[nodealign=true,colsep=2ex,rowsep=1.25ex]
        \Rnode{N1}{$\sastate_3$} &             & \Rnode{N2}{$\sastate_2$}         &  &
        \Rnode{N3}{$\sastate_1$} & \pnode{N34} & \Rnode{N4}{$\sastateset_1$}      &  &
        \Rnode{N5}{$\cdots$}     &             & \Rnode{N6}{$\sastateset_2$}      &  &
        \Rnode{N7}{$\cdots$}     &             & \Rnode{N8}{$\sastateset_3$}      &  &
        \Rnode{N12}{$\cdots$}    &             &                                  &  &
        \Rnode{N9}{$\cdots$}     &             & \Rnode{N10}{$\brac{\sastateset_\branch \cup \ldots}$} &  &
        \Rnode{N11}{$\cdots$} \\

        \psset{nodesep=.5ex,angle=-90,linearc=.2}
        \ncline{->}{N1}{N2}^{$\sopen{}$}
        \ncline{->}{N2}{N3}^{$\sopen{}$}
        \ncline{->}{N3}{N4}^{$\cha$}
        \ncbar[arm=1.5ex,nodesepA=0]{->}{N34}{N10}
        \ncline{->}{N4}{N5}^{$\cdots$}
        \ncline{->}{N5}{N6}^{$\sclose{1}$}
        \ncline{->}{N6}{N7}^{$\cdots$}
        \ncline{->}{N7}{N8}^{$\sclose{2}$}
        \ncline{->}{N8}{N12}^{$\cdots$}
        \ncline{->}{N9}{N10}\Aput{$\cdots$}
        \ncline{->}{N10}{N11}\Aput{$\cdots$}
    \end{psmatrix}
\end{center}

\subsubsection{Formal Definition of a Run}
\label{sec:formal-sa-run}

We begin by defining the set of substacks of a stack, which are intuitively all suffixes of the stack.

\begin{defi}[$\substacks{\stackw}$]
    Given an order-$\cpdsord$ stack $\stackw$, we denote by
    $\substacks{\stackw}$
    the smallest set of stacks such that
    $\stackw \in \substacks{\stackw}$
    and if
    $\ccompose{\stacku}{\opord}{\stackv} \in \substacks{\stackw}$
    for some
    $\opord \in \set{1, \ldots, \cpdsord}$
    then
    $\stackv \in \substacks{\stackw}$.
\end{defi}

A stack automaton is essentially a stack- and collapse-aware
alternating automaton, where collapse links are treated as special cases of the
alternation.
Fix a stack automaton
\[
    \saauta = \tuple{
                  \sastates_\cpdsord,\ldots,\sastates_1,
                  \alphabet,
                  \sadelta_\cpdsord,\ldots,\sadelta_1,
                  \safinals_\cpdsord,\ldots,\safinals_1
              } \ .
\]
More formally, a run of a stack automaton over an order-$\cpdsord$ stack $\stackw$ associates to each stack
$\stackv \in \substacks{\stackw}$
at most one set of states $\sastateset_\opord$ per
$\opord \in \set{1, \ldots, \cpdsord}$.
The run is accepting if the following conditions are met.
\begin{itemize}
\item
    There is an order-$\cpdsord$ stateset
    $\sastateset_\cpdsord \subseteq \sastates_\cpdsord$
    associated with
    $\sbrac{}{\cpdsord} \in \substacks{\stackw}$
    and
    $\sastateset_\cpdsord \subseteq \safinals_\cpdsord$.

\item
    If
    $\opord \in \set{1, \ldots, \cpdsord-1}$
    and
    $\brac{\ccompose{\sbrac{}{\opord}}{(\opord+1)}{\stackv}}
     \in
     \substacks{\stackw}$
    then this stack is associated with an order-$\opord$ stateset
    $\sastateset_\opord \subseteq \safinals_\opord$.

\item
    Each
    $\brac{\ccompose{\annot{\cha}{\tuple{\opord, \idxi}}}{1}{\stackv}}
     \in
     \substacks{\stackw}$
    is associated with an order-$1$ stateset
    $\sastateset_1 \subseteq \sastates_1$
    with
    $\stackv$
    associated with an order-$1$ stateset
    $\sastateset'_1 \subseteq \sastates_1$
    and
    $\apply{\cbottom{\cpdsord}{\idxi}}{
        \ccompose{\annot{\cha}{\tuple{\opord, \idxi}}}{1}{\stackv}
     }$
    associated with an order-$\opord$ stateset
    $\sastateset_\opord \subseteq \sastates_\opord$
    such that for each
    $\sastate_1 \in \sastateset_1$
    there is a transition
    $\sastate_1 \satrancol{\cha}{\sastateset_\branch} \sastateset \in \sadelta_1$
    such that
    $\sastateset_\branch \subseteq \sastateset_\opord$
    and
    $\sastateset \subseteq \sastateset'_1$.

\item
    Each
    $\ccompose{\stacku}{\opord}{\stackv} \in \substacks{\stackw}$
    with
    $\opord \in \set{2, \ldots, \cpdsord}$
    is associated with an order-$\opord$ stateset
    $\sastateset_\opord \subseteq \sastates_\opord$
    and an order-$(\opord-1)$ stateset
    $\sastateset_{\opord-1} \subseteq \sastates_{\opord-1}$
    and $\stackv$ is associated with an order-$\opord$ stateset
    $\sastateset'_\opord \subseteq \sastates_{\opord-1}$
    such that for each
    $\sastate_\opord \in \sastateset_\opord$
    there is a transition
    $\sastate_\opord \satran{\sastate_{\opord-1}} \sastateset \in \sadelta_\opord$
    such that
    $\sastate_{\opord-1} \in \sastateset_{\opord-1}$
    and
    $\sastateset \subseteq \sastateset'_\opord$.
\end{itemize}

We write
$\stackw \in \slang{\sastate}{\saauta}$
to denote that the order-$\cpdsord$ stack $\stackw$ is accepted by $\saauta$ from
$\sastate \in \sastates_\opord$ for some $\opord$.
Similarly, for
$\sastateset \subseteq \sastates_\opord$
for some $\opord$ we write
$\stackw \in \slang{\sastateset}{\saauta}$
when $\stackw$ is accepted from each
$\sastate \in \sastateset$.
Note, if $\sastateset = \emptyset$ then all stacks are accepted.

\lmcscorrection{
    \subsubsection{Properties of Stack Automata}

    We show here that stack automata membership is linear time.
    Several further results can also be shown~\cite{BCHS12}: the sets of stacks accepted
    by these automata form an effective Boolean algebra (note that complementation
    causes a blow-up in the size of the automaton); and they accept the same family
    of collapsible stacks as the automata used by Broadbent\etal~\cite{BCOS10}.  We
    omit these here for space reasons.

    We also report that our PSPACE emptiness algorithm for stack automata~\cite{BCHS12} is not correct\footnote{%
        We thank an anonymous reviewer for pointing this out.
    }.
    Indeed, we have shown that the question, given a stack automaton $\saauta$, whether there exists a collapsible pushdown stack accepted by $\saauta$, is NEXPTIME-complete~\cite{ARXIV}.
    We again omit this proof for space reasons.
    In the sequel, we will primarily be interested in membership rather than emptiness.

    \begin{prop}[Stack Automata Membership]
        Membership of order-$\cpdsord$ stack automata can be tested in linear time
        in the size of the input stack and stack automaton.
    \end{prop}
    \proof
        Take a stack $\stackw$ and let
        \[
        \saauta = \tuple{
                      \sastates_\cpdsord,\ldots,\sastates_1,
                      \alphabet,
                      \sadelta_\cpdsord,\ldots,\sadelta_1,
                      \safinals_\cpdsord, \ldots,\safinals_1
                  } \ .
        \]

        The membership algorithm iterates from the bottom (end) of the stack to the top (beginning).
        We start at the bottom of the order-$\cpdsord$ stack.
        In particular,
        with
        $\sbrac{}{\cpdsord}$
        we associate
        $\safinals_\cpdsord$
        and for all
        $\opord \in \set{2, \ldots, \cpdsord}$
        we associate with each
        $(\ccompose{\sbrac{}{\opord}}{(\opord+1)}{\stackv}) \in \substacks{\stackw}$
        the set $\safinals_\opord$.
        It will be easy to verify at all stages that if we associate a set of states
        $\sastateset$
        with a stack, then if
        $\sastate \in \sastateset$,
        then the stack from that position is in $\slang{\sastate}{\saauta}$.
        Note, with each order $\opord$ and each stack in $\substacks{\stackw}$ we associate at most one set of states.
        Hence, the algorithm can be run in linear time.
        We extend our associations until we have associated sets of states with $\stackw$, from which membership can be tested.

        In the first case assume we have
        $\brac{\ccompose{\annot{\cha}{\tuple{\opord, \idxi}}}
                       {1}
                       {\stackv}} \in \substacks{\stackw}$
        and have associated
        $\stackv$
        with an order-$1$ stateset
        $\sastateset_1 \subseteq \sastates_1$
        and
        $\apply{\cbottom{\opord}{\idxi}}
               {\ccompose{\annot{\cha}{\tuple{\opord, \idxi}}}
                         {1}
                         {\stackv}}$
        is associated with an order-$\opord$ stateset
        $\sastateset_\branch \subseteq \sastates_\opord$.
        In this case, we associate with
        $\brac{\ccompose{\annot{\cha}{\tuple{\opord, \idxi}}}
                       {1}
                       {\stackv}}$
        the order-$1$ stateset
        $\sastateset'_1 =
         \setcomp{\sastate}
                 {\sastate \satrancol{\cha}{\sastateset'_\branch} \sastateset'_1 \in \sadelta_1
                  \land
                  \sastateset'_1 \subseteq \sastateset_1
                  \land
                  \sastateset'_\branch \subseteq \sastateset_\branch}
         \subseteq
         \sastates_1$.
        Thus, any state in $\sastateset_0$ has a transition to states from which the remainder of the stack is accepted.

        The next case is when we have an order-$(\opord-1)$ stack
        $\stacku$
        and an order-$\cpdsord$ stack
        $\stackv$
        and
        $\ccompose{\stacku}{\opord}{\stackv} \in \substacks{\stackw}$
        for some
        $\opord \in \set{2, \ldots, \cpdsord}$.
        Suppose we have associated with $\stackv$ the order-$\opord$ stateset
        $\sastateset_\opord \subseteq \sastates_\opord$.
        Moreover, suppose we have associated with
        $\ccompose{\stacku}{\opord}{\stackv} \in \substacks{\stackw}$
        an order-$(\opord-1)$ stateset
        $\sastateset_{\opord-1} \subseteq \sastates_{\opord-1}$.
        In this case we also associate with
        $\ccompose{\stacku}{\opord}{\stackv}$
        the order-$\opord$ stateset
        $\sastateset'_\opord
         =
         \setcomp{\sastate}
                 {\sastate \satran{\sastate'} \sastateset \in \sadelta_{\opord+1}
                  \land
                  \sastate' \in \sastateset_{\opord-1}
                  \land
                  \sastateset \subseteq \sastateset_\opord}$.
        Thus, there is a state
        $\sastate \in \sastateset'_\opord$
        whenever there is a transition
        $\sastate \satran{\sastate'} \sastateset$
        such that the first order-$(\opord-1)$ stack is accepted from $\sastate'$ and the remainder of the stack is accepted from $\sastateset$.

        For all
        $\opord \in \set{1,\ldots,\cpdsord}$,
        once we have associated with $\stackw$ an order-$\opord$ stateset
        $\sastateset_\opord \subseteq \sastates_\opord$,
        we can test whether
        $\stackw \in \slang{\sastateset_0}{\saauta}$
        for
        $\sastateset_0 \subseteq \sastates_\opord$
        by checking whether
        $\sastateset_0 \subseteq \sastateset_\opord$.
    \qed
}{
}


%% file: emptiness.tex
\section{Emptiness of Stack Automata}
\label{sec:emptiness}

In ICALP 2012~\cite{BCHS12} we incorrectly stated that the emptiness problem
for stack automata was PSPACE-complete.
As pointed out by an anonymous reviewer, the algorithm given actually runs in
PTIME and does not correctly implement the emptiness test.
We show that the problem is, in fact, NEXPTIME-complete.

\begin{theorem}
    Let $\cpdsord \geq w$ and $\saauta$ be an order-$\cpdsord$ stack automaton.
    Testing whether there exists an order-$\cpdsord$ collapsible pushdown stack
    $\stackw$ such that
    $\stackw \in \slang{\sastate}{\saauta}$
    for a given state $\sastate$ of $\saauta$ is NEXPTIME-complete.
\end{theorem}

The above theorem is proved in Section~\ref{sec:upper} and
Section~\ref{sec:lower} below.

\subsection{Upper Bound}
\label{sec:upper}

The upper bound can be obtained quite easily.
We know from ICALP 2012 that stack automata are equivalent to the bottom-up
automata introduced by Broadbent\etal~\cite{BCOS10}.
More formally, we have the following proposition.
The complexity is apparent from the proof presented in the paper.

\begin{prop}[\cite{BCOS10}, as Proposition 4]
    For every order-$\cpdsord$ stack automaton $\saauta$ with initial state
    $\sastate$, there is a bottom-up stack automaton $\saautb$ of size
    exponential in the size of $\saauta$ with initial
    state $\sastate'$ such that $\slang{\sastate}{\saauta} =
    \slang{\sastate'}{\saautb}$.
\end{prop}

Then, from Broadbent\etal~\cite{BCOS10}, we know the emptiness problem for
bottom-up stack automata is NP-complete.
When applied to an exponentially large automaton, this gives us NEXPTIME as
required.

\begin{prop}[\cite{BCOS10}, as Proposition 2]
    Given fixed $\cpdsord \geq 2$ and some automaton $\saautb$, deciding
    whether there exists some order-$\cpdsord$ collapsible stack that it
    accepts is NP-complete.
\end{prop}

In conclusion, we have the following proposition.

\begin{prop}
    Emptiness checking of order-$\cpdsord$ stack automata is in NEXPTIME.
\end{prop}

\subsection{Lower Bound}
\label{sec:lower}

The lower bound is by reduction from a tiling problem over a
$2^\numof\times 2^\numof$
grid.
It is known that, when $\numof$ is given in unary, there is a fixed
tiling problem for which the problem in NEXPTIME-hard in the size of
$\numof$~\cite{GI09}.
We begin by recalling the definition of a tiling problem before giving the
reduction.

\subsubsection{Tiling Problems}

\begin{defi}[Tiling Problem]
    A \emph{tiling problem} is a tuple
    $\tuple{\tiles, \hrel, \vrel, \inittile, \fintile}$
    where
        $\tiles$ is a finite set of tiles,
        $\hrel \subseteq \tiles \times \tiles$ is a horizontal matching relation,
        $\vrel \subseteq \tiles \times \tiles$ is a vertical matching relation, and
        $\inittile, \fintile \in \tiles$ are initial and final tiles respectively.
\end{defi}

A solution to a tiling problem over a $\dimension$-width and
$\dimension$-height corridor is a sequence
\[
    \begin{array}{c}
        \tile^1_1 \ldots \tile^1_\dimension \\
        \tile^2_1 \ldots \tile^2_\dimension \\
        \ldots \\
        \tile^\dimension_1 \ldots \tile^\dimension_\dimension
    \end{array}
\]
where
$\tile^1_1 = \inittile$,
$\tile^\dimension_\dimension = \fintile$,
and for all
$1 \leq i < \dimension$
and
$1 \leq j \leq \dimension$
we have
$\tuple{\tile^j_i, \tile^j_{i+1}} \in \hrel$
and for all
$1 \leq i \leq \dimension$
and
$1 \leq j < \tileheight$
we have
$\tuple{\tile^j_i, \tile^{j+1}_i} \in \vrel$.
Note, the grid layout is for presentation purposes only, and the sequence
should truly be written
$\tile^1_1 \ldots \tile^1_\dimension
 \tile^2_1 \ldots \tile^2_\dimension
 \ldots
 \tile^\dimension_1 \ldots \tile^\dimension_\dimension$.
We will assume that $\inittile$ and $\fintile$ can only appear at the beginning
and end of the tiling respectively.

In the sequel we will fix a tiling problem
$\tuple{\tiles, \hrel, \vrel, \inittile, \fintile}$
such that for any $\numof$ (in unary) finding a solution to the problem over a
$\dimension$-width and $\dimension$-height corridor where
$\dimension = 2^\numof$
is NEXPTIME-hard~\cite{GI09}.

\subsubsection{Reduction to Stack Automata Emptiness}

We first describe the shape of the stacks we wish to see, given a tiling
problem as fixed above.
Then we will show how to build a stack automaton which recognises such stacks
which encode solutions to the tiling problem.
For technical convenience, given $\numof$ in unary, we define
$\dimension = 2^\numof - 1$.

\paragraph{Encoding Solutions as Stacks}

We will define an order-$2$ stack automaton that will only accept stacks of the
following form.
The notation
$\binof{\idxi}$
for
$0 \leq \idxi \leq \dimension$
is the $\numof$-bit encoding of $\idxi$, most significant bit first, using the
stack characters $\bone$ and $\bzero$.
Collapse links are drawn as arrows and only appear if necessary.
Characters without explicit links may have any valid link as they are not
needed for the encoding.
Note, the grid structure is for presentation only and the stack should be read
as a sequence of order-$1$ stacks, from left-to-right and top-to-bottom.
\[
    \ssbrac{\begin{array}{rrrr}
        & \ssbrac{ \spacer }{1} & \ssbrac{ \spacer }{1} & \ssbrac{ \spacer }{1} \\
        \\
        \ssbrac{\ \rnode{oos}{\spacer}\ \binof{0}\ \binof{0}\ \tile_{0,0}\ }{1} &
        \ssbrac{\ \rnode{ois}{\spacer}\ \binof{0}\ \binof{1}\ \tile_{0,1}\ }{1} &
        \cdots &
        \ssbrac{\ \rnode{ons}{\spacer}\ \binof{0}\ \binof{\dimension}\ \tile_{0,\dimension}\ }{1} \\
        \\
        \pnode{iod}
        \ssbrac{\ \rnode{ios}{\spacer}\ \binof{1}\ \binof{0}\ \tile_{1,0}\ }{1} &
        \pnode{iid}
        \ssbrac{\ \rnode{iis}{\spacer}\ \binof{1}\ \binof{1}\ \tile_{1,1}\ }{1} &
        \cdots &
        \pnode{ind}
        \ssbrac{\ \rnode{ins}{\spacer}\ \binof{1}\ \binof{\dimension}\ \tile_{1,\dimension}\ }{1} \\
        \\
        \rnode{dod}{\phantom{[}}\ \rnode{dos}{\phantom{\spacer}}\ \quad\vdots\qquad\qquad &
        \rnode{did}{\phantom{[}}\ \rnode{dis}{\phantom{\spacer}}\ \quad\vdots\qquad\qquad &
        \cdots &
        \rnode{dnd}{\phantom{[}}\ \rnode{dns}{\phantom{\spacer}}\ \quad\vdots\qquad\qquad \\
        \\
        \pnode{nod} \ssbrac{\ \spacer\ \binof{\dimension}\ \binof{0}\ \tile_{\dimension,0}\ }{1} &
        \pnode{nid} \ssbrac{\ \spacer\ \binof{\dimension}\ \binof{1}\ \tile_{\dimension,1}\ }{1} &
        \cdots &
        \pnode{nnd} \ssbrac{\ \spacer\ \binof{\dimension}\ \binof{\dimension}\ \tile_{\dimension,\dimension}\ }{1}
    \end{array}}{2}
\]
{
    \psset{arcangleA=240,arcangleB=280,ncurv=1.2}
    \ncarc{->}{oos}{iod}
    \ncarc{->}{ois}{iid}
    \ncarc{->}{ons}{ind}
    \ncarc{->}{ios}{dod}
    \ncarc{->}{iis}{did}
    \ncarc{->}{ins}{dnd}
    \ncarc{->}{dos}{nod}
    \ncarc{->}{dis}{nid}
    \ncarc{->}{dns}{nnd}
}
That is, the stack begins with three stacks containing only a spacer character
$\spacer$.
This is merely for technical reasons as it allows alternating transitions
later to get started.
On the next row in the diagram we have an order-$1$ stack for each tile
position on the first row of the solution.
The next row has the next row of the solution and so on.
Each order-$1$ stack encodes a position as follows.
The topmost character is another spacer $\spacer$ which contains a collapse
link.
The collapse link is order-$2$ and points to the order-$1$ stack containing the
tile vertically below.
Note, this stack is not constructible using the standard stack operations, but
the stack still meets the definition of a collapsible stack.
After the $\spacer$ there are two numbers, giving the row index and column
index respectively.
These numbers are encoded as $\numof$-bit binary numbers using the characters
$\bone$ and $\bzero$ and appearing with the most significant bit first.
For example $\binof{0}\binof{1}$ appears on the stack as
\[
    \underbrace{\bzero \ldots \bzero \bzero}_{\text{$\numof$ bits}}
    \underbrace{\bzero \ldots \bzero \bone}_{\text{$\numof$ bits}}\ .
\]
Finally, each $\tile_{\idxi, \idxj} \in \tiles$ is a tile.

\paragraph{Recognising Tiling Solutions}

We will define an order-$2$ stack automaton which only accepts valid encodings
of solutions to the tiling problem.
There are several key properties to assert.
\begin{enumerate}
\item
\label{lbl:firstprop}
\label{lbl:rightshape}
    The stack contains a sequence of order-$1$ stacks, where the topmost three stacks contains only a spacer and subsequent stacks are of the form
    \[
        \set{\spacer} \set{\bzero,\bone}^{2 \cdot \numof} \tiles \ .
    \]

\item
\label{lbl:begin}
    The fourth topmost order-$1$ stack contains
    $\binof{0}\binof{0}$.

\item
\label{lbl:end}
    The bottommost order-$1$ stack contains
    $\binof{\dimension}\binof{\dimension}$.

\item
\label{lbl:rightnumseq}
    For all but the first three and bottomost order-$1$ stacks, if the stack
    contains
    $\binof{\idxi}\binof{\idxj}$
    then the order-$1$ stack beneath it contains
    $\binof{\idxi}\binof{\idxj+1}$
    if $\idxj < \dimension$ and
    $\binof{\idxi + 1}\binof{0}$
    if $\idxj = \dimension$.

\item
\label{lbl:collapsegrid}
    For every order-$1$ stack containing
    $\binof{\idxi}\binof{\idxj}$
    with $\idxi < \dimension$ the collapse link from $\spacer$ in the stack leads to an order-$1$ stack containing
    $\binof{\idxi+1}\binof{\idxj}$.

\item
\label{lbl:init}
    We have
    $\tile_{0,0} = \inittile$.

\item
\label{lbl:fin}
    We have
    $\tile_{\dimension,\dimension} = \fintile$.

\item
\label{lbl:horiz}
    For every
    $0 \leq \idxi \leq \dimension$
    and
    $0 \leq \idxj < \dimension$
    we have
    $\tuple{\tile_{\idxi, \idxj}, \tile_{\idxi, \idxj+1}} \in \hrel$.

\item
\label{lbl:lastprop}
\label{lbl:vert}
    For every
    $0 \leq \idxi < \dimension$
    and
    $0 \leq \idxj \leq \dimension$
    we have
    $\tuple{\tile_{\idxi, \idxj}, \tile_{\idxi+1, \idxj}} \in \vrel$.
\end{enumerate}
There is a stack satisfying the above properties iff there is a solution to the
tiling problem.
At this point, the experienced reader may see how alternating automata can be
used to enforce the above properties, as the manipulation of short binary
numbers is fairly standard.

Note, if we did not assert Property~\ref{lbl:rightnumseq} we could have
multiple stacks each with the same index
$\binof{\idxi}\binof{\idxj}$.
If this were the case then Property~\ref{lbl:collapsegrid} would not ensure
that the collapse links encoded a grid.
Thus, our encoding will crucially rely on the width and height being fixed.
That is, our encoding will not extend to EXPSPACE Turing machines which may
have an unbounded corridor height.

We define an order-$2$ stack automaton recognising only such stacks.
In particular, we have
\[
        \saauta = \tuple{
                      \sastates_2, \sastates_1,
                      \alphabet,
                      \sadelta_2,\sadelta_1,
                      \safinals_2,\safinals_1
                  }
\]
where the alphabet, states, and transitions are defined during the description
below.
We simultaneously argue for the correctness of the definition.

\begin{prop}
    The tiling problem
    $\tuple{\tiles, \hrel, \vrel, \inittile, \fintile}$
    has a solution over a
    $2^\numof \times 2^\numof$
    corridor iff the order-$2$ stack automaton $\saauta$ is non-empty.
\end{prop}

\newcommand\qfin[1]{q^{#1}_f}
\newcommand\qinit{q_I}
\newcommand\qspacer{q_\spacer}
\newcommand\qprop[1]{p_{#1}}
\newcommand\qpropshape{\qprop{\ref{lbl:rightshape}}}
\newcommand\qpropbegin{\qprop{\ref{lbl:begin}}}
\newcommand\qpropend{\qprop{\ref{lbl:end}}}
\newcommand\qpropseq{\qprop{\ref{lbl:rightnumseq}}}
\newcommand\qpropgrid{\qprop{\ref{lbl:collapsegrid}}}
\newcommand\qpropinit{\qprop{\ref{lbl:init}}}
\newcommand\qpropfin{\qprop{\ref{lbl:fin}}}
\newcommand\qproph{\qprop{\ref{lbl:horiz}}}
\newcommand\qpropv{\qprop{\ref{lbl:vert}}}
\newcommand\qpropfirst{\qprop{\ref{lbl:firstprop}}}
\newcommand\qproplast{\qprop{\ref{lbl:lastprop}}}
\newcommand\qbitzero[1]{q2^0_{#1}}
\newcommand\qbitone[1]{q2^1_{#1}}
\newcommand\qbitzeron[1]{q'2^0_{#1}}
\newcommand\qbitonen[1]{q'2^1_{#1}}
\newcommand\qbitzeroi[1]{q1^0_{#1}}
\newcommand\qbitonei[1]{q1^1_{#1}}
\newcommand\qbitoneil[1]{q1^1_{#1,l}}
\newcommand\qbitzeroni[1]{q'1^0_{#1}}
\newcommand\qbitoneni[1]{q'1^1_{#1}}
\newcommand\qany{q_\ast}
\newcommand\qshapeone{q_S}
\newcommand\qnbits[1]{q^{#1}_B}
\newcommand\qeqbits[1]{q^{#1}_{=}}
\newcommand\qeqbitsis[2]{q^{#1,#2}_{=}}
\newcommand\qeqbitsisn[2]{q^{#1,#2}_{=,\$}}
\newcommand\qeqbitsl[1]{q^{#1}_{=,l}}
\newcommand\qeqbitsisl[2]{q^{#1,#2}_{=,l}}
\newcommand\qeqbitsisll[2]{q^{#1,#2}_{=,l,l}}
\newcommand\qeqbitsisnl[2]{q^{#1,#2}_{=,l,\$}}
\newcommand\qzerozero{q_{\binof{0}\binof{0}}}
\newcommand\qnn{q_{\binof{\dimension}\binof{\dimension}}}
\newcommand\qrn{q_{\binof{\dimension}\binof{\ast}}}
\newcommand\qcn{q_{\binof{\ast}\binof{\dimension}}}
\newcommand\qrightz[1]{z_{#1}}
\newcommand\qrightone[1]{o_{#1}}
\newcommand\qrightzi[1]{z'_{#1}}
\newcommand\qrightonei[1]{o'_{#1}}
\newcommand\qbitis[2]{b^{#2}_{#1}}
\newcommand\qpass[1]{\left[#1\right]_\spacer}
\newcommand\qhastile[1]{q_{#1}}
\newcommand\qhtiles[2]{q^h_{\tuple{#1,#2}}}
\newcommand\qhtile[1]{q^h_{#1}}
\newcommand\qvtile[1]{q^v_{#1}}
\newcommand\qvtilelink[1]{q^\spacer_{#1}}
\newcommand\qvtilenext[1]{q^{v'}_{#1}}

The alphabet is
$\alphabet = \set{\spacer, \bzero, \bone} \cup \tiles$.
The only accepting states are
\[
    \safinals_2 = \set{\qfin{2}} \subset \sastates_2
    \qquad
    \text{and}
    \qquad
    \safinals_1 = \set{\qfin{1}} \subset \sastates_1\ \ .
\]

The initial state is
$\qinit \in \sastates_2$
and there is a single transition from it which leads to a state for each
property above:
\[
    \qinit
    \satran{\qspacer}
    \set{\qpropfirst, \ldots, \qproplast}
    \in
    \sadelta_2
\]
where
$\qspacer \in \sastates_1$
and
$\qpropfirst, \ldots, \qproplast
 \in
 \sastates_2$.
From $\qspacer$ we simply check the stack contains only a spacer:
\[
    \qspacer \satrancol{\spacer}{\emptyset} \set{\qfin{1}} \ .
\]
We also have a state
$\qany \in \sastates_1$
that accepts any order-$1$ stack.
From it there is a transition
\[
    \qany \satrancol{\spacer}{\emptyset} \emptyset \in \sadelta_1 \ .
\]

The remaining properties are more involved.
\begin{enumerate}
\item
    To check that the stack is of the right basic shape, we have
    $\qpropshape^1, \qpropshape^2 \in \sastates_2$
    with
    \[
        \qpropshape \satran{\qspacer} \set{\qpropshape^1},
        \qpropshape^1 \satran{\qspacer} \set{\qpropshape^2}
        \in
        \sadelta_2
    \]
    to recognise the leading spacer stacks, and
    \[
        \qpropshape^2 \satran{\qshapeone} \set{\qpropshape^2},
        \qpropshape^2 \satran{\qshapeone} \set{\qfin{2}}
        \in
        \sadelta_2
    \]
    to recognise the remaining stacks, where $\qshapeone$ is a state for
    asserting the correct shape of order-$1$ stacks.
    Observe that the transition to $\qfin{2}$ can only be and must be used to
    read the bottommost order-$1$ stack in any accepting run.

    In particular, we have
    $\qshapeone,
     \qnbits{0}, \ldots, \qnbits{2\cdot\numof}
     \in
     \sastates_1$
    where we first assert the leading spacer
    \[
        \qshapeone \satrancol{\spacer}{\emptyset} \set{\qnbits{2\cdot\numof}}
        \in
        \sadelta_1
    \]
    and then that we have $2\cdot\numof$ bits with
    \[
        \qnbits{\idxi} \satrancol{\bzero}{\emptyset} \set{\qnbits{\idxi-1}},
        \qnbits{\idxi} \satrancol{\bone}{\emptyset} \set{\qnbits{\idxi-1}}
        \in
        \sadelta_1
    \]
    for all
    $1 \leq \idxi \leq 2\cdot\numof$
    and finally a trailing tile with
    \[
        \qnbits{0} \satrancol{\tile}{\emptyset} \set{\qfin{1}}
        \in
        \sadelta_1
    \]
    for all
    $\tile \in \tiles$.

\item
    To check the topmost non-spacer stack contains
    $\binof{0}\binof{0}$
    we have
    $\qpropbegin, \qpropbegin^1, \qpropbegin^2
     \in
     \sastates_2$
    and
    \[
        \qpropbegin \satran{\qany} \set{\qpropbegin^1},
        \qpropbegin^1 \satran{\qany} \set{\qpropbegin^2},
        \qpropbegin^2 \satran{\qzerozero} \set{\qfin{2}}
        \in
        \sadelta_2
    \]
    where, to check the order-$1$ stack we have
    $\qzerozero, \qzerozero^1, \ldots, \qzerozero^{2\cdot\numof}
     \in
     \sastates_1$
    and, for all
    $2 \leq \idxi \leq 2\cdot\numof$,
    \[
        \qzerozero
            \satrancol{\spacer}{\emptyset} \set{\qzerozero^{2\cdot\numof}},
        \qzerozero^{\idxi}
            \satrancol{\bzero}{\emptyset} \set{\qzerozero^{\idxi-1}},
        \qzerozero^1
            \satrancol{\bzero}{\emptyset} \emptyset
        \in
        \sadelta_1 \ .
    \]

\item
    To check the bottommost order-$1$ stack contains
    $\binof{\dimension}\binof{\dimension}$
    we have
    $\qpropend \in \sastates_2$
    and
    \[
        \qpropend \satran{\qany} \set{\qpropend},
        \qpropend \satran{\qnn} \set{\qfin{2}}
        \in
        \sadelta_2
    \]
    where, to check the order-$1$ stack we have
    $\qnn, \qnn^1, \ldots, \qnn^{2\cdot\numof}
     \in
     \sastates_1$
    and, for all
    $2 \leq \idxi \leq 2\cdot\numof$,
    \[
        \qnn
            \satrancol{\spacer}{\emptyset} \set{\qnn^{2\cdot\numof}},
        \qnn^{\idxi}
            \satrancol{\bone}{\emptyset} \set{\qnn^{\idxi-1}},
        \qnn^1
            \satrancol{\bone}{\emptyset} \emptyset
        \in
        \sadelta_1 \ .
    \]

\item
    To check the sequence of binary numbers runs from
    $\binof{0}\binof{0}$
    to
    $\binof{\dimension}\binof{\dimension}$
    first notice that we can consider
    $\binof{\idxi}\binof{\idxj}$
    as a $(2 \cdot \numof)$-bit binary number.
    In this case each stack contains the successor of the stack above it.

    We will make use of the following auxilliary states in $\sastates_2$.
    That is, for all
    $1 \leq \idxi \leq 2 \cdot \numof$
    we have
    $\qbitzero{\idxi}, \qbitone{\idxi} \in \sastates_2$.
    Note, the choice of notation here indicates that we are dealing with $(2
    \cdot \numof)$-bit numbers instead of $\numof$-bit numbers.
    We will need similar states for $\numof$-bit numbers later.

    Intuitively, these states are used to check that binary encodings succeed
    each other.
    Take the two numbers $10011$ and $10100$.
    We will number bits from right to left, starting at $1$.
    Hence, in the first number, the first and second bits are $1$, the third is
    $0$, the fourth is $0$, and the fifth is $1$.
    Note the latter binary number is the increment of the former.
    This can be identified by noticing that the rightmost $0$ is bit $3$.
    In the incremented number, bit $3$ is now a $1$ and all bits to the right
    are now $0$.
    Thus, $\qbitzero{\idxi}$ will assert that the $\idxi$th bit must be the
    rigthmost $0$, and $\qbitone{\idxi}$ will assert that the rightmost $1$
    must be the $\idxi$th bit.
    Now, if one order-$1$ stack appears directly below the other, the upper
    stack will be accepted by some $\qbitzero{\idxi}$ and the lower stack will
    be accepted from the corresponding $\qbitone{\idxi}$.

    We will also need to make use of states
    $\qeqbits{1},\ldots,\qeqbits{2\cdot\numof} \in \sastates_2$
    to assert the equality of bits at each position in the binary numbers
    between the topmost two stacks.
    From these we will guess what the value of the bits are by moving to states
    $\qeqbitsis{\idxi}{\bzero},\qeqbitsis{\idxi}{\bone} \in \sastates_2$
    for all
    $1 \leq \idxi \leq 2\cdot\numof$.

    The reader may notice here that there is some delay in setting up the right
    states.
    For example, to check the equality of bits $1$ and $2$, the automaton first
    needs to read a stack and move to $\qeqbits{1}$ and $\qeqbits{2}$, and then
    needs to guess the bit values (say $\bzero$) and then, while reading
    another stack, move to
    $\qeqbitsis{1}{\bzero}$ and $\qeqbitsis{2}{\bzero}$.
    This is why we have leading stacks only containing a spacer, to allow the
    look ahead to wind up.
    Thus, as well as checking the immediate stacks, there will be some states
    for checking the stacks ahead.

    Thus, for each
    $1 \leq \idxi \leq 2\cdot\numof$
    we have from $\qpropseq$ the transition
    \[
        \qpropseq
        \satran{\qany}
        \set{\qpropseq,
             \qbitzero{\idxi},
             \qeqbits{\idxi+1},
             \ldots,
             \qeqbits{2\cdot\numof}}
        \in
        \sadelta_2 \ .
    \]
    Notice, there are only polynomially many such transitions.
    The state $\qpropseq$ appears to assert the sequence property of the next
    stack.
    Next, we make another set of transitions to guess the value of the bits to be
    tested by
    $\qeqbits{\idxi}$.
    For this we have, for all
    $1 \leq \idxi \leq 2\cdot\numof$,
    \[
        \qbitzero{\idxi} \satran{\qany} \set{\qbitzeron{\idxi}},
        \qeqbits{\idxi} \satran{\qany} \set{\qeqbitsis{\idxi}{\bzero}},
        \qeqbits{\idxi} \satran{\qany} \set{\qeqbitsis{\idxi}{\bone}}
        \in
        \sadelta_2
    \]
    with
    $\qbitzeron{\idxi} \in \sastates_2$.

    We are now ready to begin testing that the stacks are in sequence.
    That is, we check the current and succeeding stack all agree on the bits
    asserted by the states
    $\qeqbitsis{\idxi}{\bzero}$ and $\qeqbitsis{\idxi}{\bone}$
    and the increment of the rightmost $\bzero$ is done correctly.
    For this we have the transitions
    \[
        \qbitzeron{\idxi}
            \satran{\qrightz{\idxi}} \set{\qbitone{\idxi}},
        \qeqbitsis{\idxi}{\bzero}
            \satran{\qbitis{\idxi}{\bzero}} \set{\qeqbitsisn{\idxi}{\bzero}},
        \qeqbitsis{\idxi}{\bone}
            \satran{\qbitis{\idxi}{\bone}} \set{\qeqbitsisn{\idxi}{\bone}}
        \in
        \sadelta_2
    \]
    for all
    $1 \leq \idxi \leq 2\cdot\numof$
    with
    $\qeqbitsisn{\idxi}{\bzero}, \qeqbitsisn{\idxi}{\bone} \in \sastates_2$.
    We delay the description of the order-$1$ states until we are finished at
    order-$2$ (see below).
    To complete the treatment at order-$2$ we have transitions that check the
    successor stack.
    These are, for all
    $1 \leq \idxi \leq 2\cdot\numof$,
    \[
        \qbitone{\idxi}
            \satran{\qrightone{\idxi}} \emptyset,
        \qeqbitsisn{\idxi}{\bzero}
            \satran{\qbitis{\idxi}{\bzero}} \emptyset,
        \qeqbitsisn{\idxi}{\bone}
            \satran{\qbitis{\idxi}{\bone}} \emptyset
        \in
        \sadelta_2 \ .
    \]
    We also need transitions which allow the automaton to leave the bottommost
    stack unchecked (it is checked in the case above).
    For this we allow all states that should check a stack to simply move to
    the final state.
    That is,
    \[
        \qpropseq
            \satran{\qany} \set{\qfin{2}},
        \qbitzero{\idxi}
            \satran{\qany} \set{\qfin{2}},
        \qbitzeron{\idxi}
            \satran{\qany} \set{\qfin{2}},
        \qeqbits{\idxi}
            \satran{\qany} \set{\qfin{2}},
        \qeqbitsis{\idxi}{\bzero}
            \satran{\qany} \set{\qfin{2}},
        \qeqbitsis{\idxi}{\bone}
            \satran{\qany} \set{\qfin{2}}
        \in
        \sadelta_2
    \]
    for all
    $1 \leq \idxi \leq 2\cdot\numof$.

    There are several properties the above transitions need to assert at
    order-$1$.
    First we have the states
    $\qrightz{\idxi}, \qrightz{\idxi,\idxj} \in \sastates_1$
    for all
    $1 \leq \idxi, \idxj \leq 2\cdot\numof$.
    These states check the rightmost $\bzero$ appears at position $\idxi$.
    We first skip the leading spacer with
    \[
        \qrightz{\idxi}
            \satrancol{\spacer}{\emptyset} \set{\qrightz{\idxi,2\cdot\numof}}
        \in
        \sadelta_1 \ .
    \]
    Then we have, for all
    $1 \leq \idxi < \idxj \leq 2\cdot\numof$,
    \[
        \qrightz{\idxi,\idxj}
            \satrancol{\bzero}{\emptyset} \set{\qrightz{\idxi,\idxj-1}},
        \qrightz{\idxi,\idxj}
            \satrancol{\bone}{\emptyset} \set{\qrightz{\idxi,\idxj-1}},
        \in
        \sadelta_1
    \]
    and for all
    $1 < \idxi \leq 2\cdot\numof$
    \[
        \qrightz{\idxi,\idxi}
            \satrancol{\bzero}{\emptyset} \set{\qrightz{\idxi, \idxi-1}},
        \qrightz{1,1}
            \satrancol{\bzero}{\emptyset} \emptyset
        \in
        \sadelta_1
    \]
    and for all
    $2 \leq \idxj < \idxi \leq 2\cdot\numof$
    \[
        \qrightz{\idxi,\idxj}
            \satrancol{\bone}{\emptyset} \set{\qrightz{\idxi, \idxj-1}},
        \qrightz{\idxi,1}
            \satrancol{\bone}{\emptyset} \emptyset,
        \qrightz{2,1}
            \satrancol{\bone}{\emptyset} \emptyset
        \in
        \sadelta_1 \ .
    \]

    Similarly, we check the rightmost $\bone$ bit with
    $\qrightone{\idxi}, \qrightone{\idxi,\idxj} \in \sastates_1$
    for all
    $1 \leq \idxi, \idxj \leq 2\cdot\numof$.
    We first skip the leading spacer with
    \[
        \qrightone{\idxi}
            \satrancol{\spacer}{\emptyset} \set{\qrightone{\idxi,2\cdot\numof}}
        \in
        \sadelta_1 \ .
    \]
    Then we have, for all
    $1 \leq \idxi < \idxj \leq 2\cdot\numof$,
    \[
        \qrightone{\idxi,\idxj}
            \satrancol{\bzero}{\emptyset} \set{\qrightone{\idxi,\idxj-1}},
        \qrightone{\idxi,\idxj}
            \satrancol{\bone}{\emptyset} \set{\qrightone{\idxi,\idxj-1}},
        \in
        \sadelta_1
    \]
    and for all
    $1 < \idxi \leq 2\cdot\numof$
    \[
        \qrightone{\idxi,\idxi}
            \satrancol{\bone}{\emptyset} \set{\qrightone{\idxi, \idxi-1}},
        \qrightone{1,1}
            \satrancol{\bone}{\emptyset} \emptyset
        \in
        \sadelta_1
    \]
    and for all
    $2 \leq \idxj < \idxi \leq 2\cdot\numof$
    \[
        \qrightone{\idxi,\idxj}
            \satrancol{\bzero}{\emptyset} \set{\qrightone{\idxi, \idxj-1}},
        \qrightone{\idxi,1}
            \satrancol{\bzero}{\emptyset} \emptyset,
        \qrightone{2,1}
            \satrancol{\bzero}{\emptyset} \emptyset
        \in
        \sadelta_1 \ .
    \]

    Next, we check that the $\idxi$th bit is a $\bzero$ with the states
    $\qbitis{\idxi}{\bzero}, \qbitis{\idxi,\idxj}{\bzero} \in \sastates_1$
    for all
    $1 \leq \idxi \leq \idxj \leq 2\cdot\numof$.
    We first skip the leading spacer with
    \[
        \qbitis{\idxi}{\bzero}
            \satrancol{\spacer}{\emptyset}
            \set{\qbitis{\idxi,2\cdot\numof}{\bzero}}
        \in
        \sadelta_1 \ .
    \]
    Then we have, for all
    $1 \leq \idxi < \idxj \leq 2\cdot\numof$,
    \[
        \qbitis{\idxi,\idxj}{\bzero}
            \satrancol{\bzero}{\emptyset} \set{\qbitis{\idxi,\idxj-1}{\bzero}},
        \qbitis{\idxi,\idxj}{\bzero}
            \satrancol{\bone}{\emptyset} \set{\qbitis{\idxi,\idxj-1}{\bzero}},
        \in
        \sadelta_1
    \]
    and for all
    $1 \leq \idxi \leq 2\cdot\numof$
    \[
        \qbitis{\idxi,\idxi}{\bzero}
            \satrancol{\bzero}{\emptyset} \emptyset
        \in
        \sadelta_1 \ .
    \]
    Similarly for $\bone$ and $\qbitis{\idxi}{\bone}$.

\item
    To check that the collapse links form a grid, we follow a similar strategy
    to the previous case, with some differences.
    Aside from the change in the handling of successors, the main change is
    that instead of checking adjacent order-$1$ stacks, the pairs of stacks we
    check are separated by collapse links.
    Thus we have to be able to send order-$2$ states through the collapse links
    from where they can assert properties of the linked-to order-$1$ stack.
    For this we introduce states of the form
    $\qpass{\sastate} \in \sastates_1$
    for all
    $\sastate \in \sastates_2$.
    From these states we have only the transition
    \[
        \qpass{\sastate} \satrancol{\spacer}{\set{\sastate}} \emptyset \ .
    \]

    Recall, when following collapse links we need to move from a stack containing
    $\binof{\idxi}\binof{\idxj}$
    to
    $\binof{\idxi+1}\binof{\idxj}$.
    We use a slight modification of the previous strategy for testing the
    increment of $\idxi$, except we restrict the checks to the leftmost
    $\numof$ bits.
    The equality checks are almost the same, except we have to move through the
    collapse links instead of down the stack.

    Thus, similar to $\qbitzero{\idxi}$ and $\qbitone{\idxi}$ we use for all
    $\numof < \idxi \leq 2\cdot\numof$
    the states
    $\qbitzeroi{\idxi}, \qbitonei{\idxi} \in \sastates_2$
    to check the rightmost $\bzero$ or $\bone$ respectively in the leftmost
    $\numof$ bits appears at position $\idxi$.
    We introduce as well
    $\qbitoneil{\idxi} \in \sastates_2$
    to pass $\qbitonei{\idxi}$ through the link.
    We also have
    $\qeqbitsl{\idxi} \in \sastates_2$
    for all
    $1 \leq \idxi \leq 2\cdot\numof$
    to verify the equality of bit positions through links.

    As in the previous case, we wind up the alternation with the following
    transitions
    \[
        \qpropgrid
        \satran{\qany}
        \set{\qpropgrid,
             \qeqbitsl{1},
             \ldots
             \qeqbitsl{\numof},
             \qbitzeroi{\idxi},
             \qeqbitsl{\idxi+1},
             \ldots,
             \qeqbitsl{2\cdot\numof}}
        \in
        \sadelta_2
    \]
    for all
    $\numof < \idxi \leq 2\cdot\numof$
    Notice again there are only a polynomial number of such transitions.
    Here, $\qpropgrid$ appears on the right to verify the property at the next
    grid position (hence we verify all positions).
    We check the equality of the first $\numof$ bits since these encode the
    column index which has to be equal.
    The remaining states verify that the row position is incremented correctly
    through the links.

    Next, we take one more step to guess the values of the bit positions being
    tested for equality.
    Thus we have transitions
    \[
        \qeqbitsl{\idxi}
            \satran{\qany} \set{\qeqbitsisl{\idxi}{\bzero},
                                \qeqbitsisll{\idxi}{\bzero}},
        \qeqbitsl{\idxi}
            \satran{\qany} \set{\qeqbitsisl{\idxi}{\bone},
                                \qeqbitsisll{\idxi}{\bone}}
        \in
        \sadelta_2
    \]
    for all
    $1 \leq \idxi \leq 2\cdot\numof$
    where
    $\qeqbitsisl{\idxi}{\bzero},
     \qeqbitsisll{\idxi}{\bzero},
     \qeqbitsisl{\idxi}{\bone},
     \qeqbitsisll{\idxi}{\bone}
     \in
     \sastates_2$.
    Note, the states like
    $\qeqbitsisll{\idxi}{\bzero}$
    are used to pass the equals check through the link as will be seen below.
    At the same time we fire
    \[
        \qbitzeroi{\idxi}
            \satran{\qany} \set{\qbitzeroni{\idxi}, \qbitoneil{\idxi}}
        \in
        \sadelta_2
    \]
    where
    $\qbitzeroni{\idxi} \in \sastates_2$.

    Now we do the checking.
    To do the bit equality checks we have similar transitions to previously.
    That is for all
    $1 \leq \idxi \leq 2\cdot\numof$
    \[
        \qeqbitsisl{\idxi}{\bzero} \satran{\qbitis{\idxi}{\bzero}} \emptyset,
        \qeqbitsisl{\idxi}{\bone} \satran{\qbitis{\idxi}{\bone}} \emptyset
        \in
        \sadelta_2
    \]
    and to pass the checks through to the linked stacks we have
    \[
        \qeqbitsisll{\idxi}{\bzero}
            \satran{\qpass{\qeqbitsisnl{\idxi}{\bzero}}} \emptyset,
        \qeqbitsisll{\idxi}{\bone}
            \satran{\qpass{\qeqbitsisnl{\idxi}{\bone}}} \emptyset
        \in
        \sadelta_2
    \]
    and once the states have been passed through the links we have
    \[
        \qeqbitsisnl{\idxi}{\bzero}
            \satran{\qbitis{\idxi}{\bzero}} \emptyset,
        \qeqbitsisnl{\idxi}{\bone}
            \satran{\qbitis{\idxi}{\bone}} \emptyset
        \in
        \sadelta_2 \ .
    \]
    Next, to check the increment we need for each
    $\numof < \idxi \leq 2\cdot\numof$
    \[
        \qbitzeroni{\idxi}
            \satran{\qrightzi{\idxi}} \emptyset,
        \qbitoneil{\idxi}
            \satran{\qpass{\qbitonei{\idxi}}} \emptyset
        \in
        \sadelta_2
    \]
    and once the state has been passed through the link we have
    \[
        \qbitonei{\idxi}
            \satran{\qrightonei{\idxi}} \emptyset
        \in
        \sadelta_2
    \]
    where the order-$1$ states are described after we complete our description
    of the order-$2$ transitions.

    To complete the order-$2$ description we need a way to handle the final row
    of the grid.
    For this we use a state
    $\qrn \in \sastates_1$
    which asserts the part of the binary number encoding the row consists of
    only $\bone$ characters.
    We give its transitions below.
    Thus, the above states can avoid checking a stack by asserting it appears
    on the final row.
    That is
    \begin{multline*}
        \qpropgrid
            \satran{\qrn} \emptyset,
        \qbitzeroi{\idxi}
            \satran{\qrn} \emptyset,
        \qbitzeroni{\idxi}
            \satran{\qrn} \emptyset, \\
        \qeqbitsl{\idxi}
            \satran{\qrn} \emptyset,
        \qeqbitsisl{\idxi}{\bzero}
            \satran{\qrn} \emptyset,
        \qeqbitsisl{\idxi}{\bone}
            \satran{\qrn} \emptyset,
        \qeqbitsisll{\idxi}{\bzero}
            \satran{\qrn} \emptyset,
        \qeqbitsisll{\idxi}{\bone}
            \satran{\qrn} \emptyset
        \in
        \sadelta_2
    \end{multline*}
    for all
    $1 \leq \idxi \leq 2\cdot\numof$.

    It remains to check several properties the above transitions need to assert
    at order-$1$.
    We have the states
    $\qrightzi{\idxi}, \qrightzi{\idxi,\idxj} \in \sastates_1$
    for all
    $\numof < \idxi, \idxj \leq 2\cdot\numof$.
    These states check the rightmost $\bzero$ appears at position $\idxi$ in
    the encoding of the row number.
    We first skip the leading spacer with
    \[
        \qrightzi{\idxi}
            \satrancol{\spacer}{\emptyset} \set{\qrightzi{\idxi,2\cdot\numof}}
        \in
        \sadelta_1 \ .
    \]
    Then we have, for all
    $\numof < \idxi < \idxj \leq 2\cdot\numof$,
    \[
        \qrightzi{\idxi,\idxj}
            \satrancol{\bzero}{\emptyset} \set{\qrightzi{\idxi,\idxj-1}},
        \qrightzi{\idxi,\idxj}
            \satrancol{\bone}{\emptyset} \set{\qrightzi{\idxi,\idxj-1}},
        \in
        \sadelta_1
    \]
    and for all
    $\numof + 1 < \idxi \leq 2\cdot\numof$
    \[
        \qrightzi{\idxi,\idxi}
            \satrancol{\bzero}{\emptyset} \set{\qrightzi{\idxi, \idxi-1}},
        \qrightzi{\numof+1,\numof+1}
            \satrancol{\bzero}{\emptyset} \emptyset
        \in
        \sadelta_1
    \]
    and for all
    $\numof + 1 < \idxj < \idxi \leq 2\cdot\numof$
    \[
        \qrightzi{\idxi,\idxj}
            \satrancol{\bone}{\emptyset} \set{\qrightzi{\idxi, \idxj-1}},
        \qrightzi{\idxi,\numof + 1}
            \satrancol{\bone}{\emptyset} \emptyset,
        \qrightz{\numof + 2,\numof + 1}
            \satrancol{\bone}{\emptyset} \emptyset
        \in
        \sadelta_1 \ .
    \]

    Similarly, we check the rightmost $\bone$ bit with
    $\qrightonei{\idxi}, \qrightonei{\idxi,\idxj} \in \sastates_1$
    for all
    $\numof < \idxi, \idxj \leq 2\cdot\numof$.
    We first skip the leading spacer with
    \[
        \qrightonei{\idxi}
            \satrancol{\spacer}{\emptyset} \set{\qrightonei{\idxi,2\cdot\numof}}
        \in
        \sadelta_1 \ .
    \]
    Then we have, for all
    $\numof < \idxi < \idxj \leq 2\cdot\numof$,
    \[
        \qrightonei{\idxi,\idxj}
            \satrancol{\bzero}{\emptyset} \set{\qrightonei{\idxi,\idxj-1}},
        \qrightonei{\idxi,\idxj}
            \satrancol{\bone}{\emptyset} \set{\qrightonei{\idxi,\idxj-1}},
        \in
        \sadelta_1
    \]
    and for all
    $\numof + 1 < \idxi \leq 2\cdot\numof$
    \[
        \qrightonei{\idxi,\idxi}
            \satrancol{\bone}{\emptyset} \set{\qrightone{\idxi, \idxi-1}},
        \qrightonei{\numof + 1, \numof + 1}
            \satrancol{\bone}{\emptyset} \emptyset
        \in
        \sadelta_1
    \]
    and for all
    $\numof + 2 < \idxj < \idxi \leq 2\cdot\numof$
    \[
        \qrightonei{\idxi,\idxj}
            \satrancol{\bzero}{\emptyset} \set{\qrightonei{\idxi, \idxj-1}},
        \qrightonei{\idxi,\numof + 1}
            \satrancol{\bzero}{\emptyset} \emptyset,
        \qrightone{\numof + 2, \numof + 1}
            \satrancol{\bzero}{\emptyset} \emptyset
        \in
        \sadelta_1 \ .
    \]

    It only remains to define the transitions from $\qrn$.
    For this we introduce several intermediate states
    $\qrn, \qrn^{\numof + 1}, \ldots, \qrn^{2\cdot\numof}
     \in \sastates_1$
    and the transitions
    \[
        \qrn \satrancol{\spacer}{\emptyset} \set{\qrn^{2\cdot\numof}},
        \qrn^{2\cdot\numof}
            \satrancol{\bone}{\emptyset} \set{\qrn^{2\cdot\numof - 1}},
        \ldots,
        \qrn^{\numof + 2}
            \satrancol{\bone}{\emptyset} \set{\qrn^{\numof + 1}},
        \qrn^{\numof + 1}
            \satrancol{\bone}{\emptyset} \emptyset
        \in
        \sadelta_1 \ .
    \]

\item
    Next we check that the first tile is $\inittile$.
    We will first introduce some order-$1$ states for checking the tile in a
    stack.
    That is, for each
    $\tile \in \tiles$
    we have
    $\qhastile{\tile} \in \sastates_1$
    from which we have the transitions
    \[
        \qhastile{\tile} \satrancol{\cha}{\emptyset} \set{\qhastile{\tile}},
        \qhastile{\tile} \satrancol{\tile}{\emptyset} \emptyset
        \in
        \sadelta_1
    \]
    where $\cha$ ranges over
    $\alphabet \setminus \tiles$.

    Then to check the first tile we use
    \[
        \qpropinit \satran{\qany} \set{\qpropinit^1},
        \qpropinit \satran{\qany} \set{\qpropinit^2},
        \qpropinit \satran{\qhastile{\inittile}} \emptyset
        \in
        \sadelta_2
    \]
    with the states
    $\qpropinit, \qpropinit^1, \qpropinit^2 \in \sastates_2$.
    These transitions simply skip the leading spacer stacks and check the first
    stack encoding a cell holds $\inittile$.

\item
    To check the final tile is $\fintile$ we use
    \[
        \qpropfin \satran{\qany} \set{\qpropfin},
        \qpropfin \satran{\qhastile{\fintile}} \set{\qfin{2}}
        \in
        \sadelta_2
    \]
    which simply iterate to the final order-$1$ stack and verify it contains
    $\fintile$.

\item
    Next we need to define transitions that check the horizonal tiling relation.
    We will begin with a transition reading the first spacer
    \[
        \qproph \satran{\qany} \set{\qproph'} \in \sadelta_2
    \]
    where
    $\qproph, \qproph' \in \sastates_2$.

    Next we will guess the pair
    $\tuple{\tile, \tile'} \in \hrel$
    that the following two order-$1$ stacks will contain.
    For this we will need order-$2$ states
    $\qhtiles{\tile}{\tile'} \in \sastates_2$
    for each
    $\tuple{\tile, \tile'} \in \hrel$
    from which we verify the following two order-$1$ stacks contain $\tile$ and
    $\tile'$ respectively.
    In other words, the horizontal tiling relation is satisfied.
    The transitions we have are
    \[
        \qproph' \satran{\qany} \set{\qproph', \qhtiles{\tile}{\tile'}}
        \in
        \sadelta_2
    \]
    where $\qproph'$ appears on the right hand side to assert the horizontal
    tiling relation over subsequent pairs of order-$1$ stacks.
    From each
    $\qhtiles{\tile}{\tile'}$
    we have
    \[
        \qhtiles{\tile}{\tile'} \satran{\qhastile{\tile}} \set{\qhtile{\tile'}},
        \qhtile{\tile'} \satran{\qhastile{\tile'}} \emptyset
        \in
        \sadelta_2
    \]
    where
    $\qhtile{\tile'} \in \sastates_2$
    are intermediate states for each required $\tile'$.

    Since the horizontal tiling does not apply to the final tile in each row,
    we have transitions allowing the condition to be relaxed here.
    That is, we have, for each
    $\tuple{\tile, \tile'} \in \hrel$,
    \[
        \qproph' \satran{\qcn} \set{\qproph'},
        \qproph' \satran{\qcn} \set{\qfin{2}},
        \qhtiles{\tile}{\tile'} \satran{\qcn} \emptyset
        \in
        \sadelta_2 \ .
    \]
    The transitions from $\qproph'$ simply keep passing the state verifying the
    tiling relation along, or terminates if the bottommost stack is read.
    From
    $\qhtiles{\tile}{\tile'}$
    we simply dismiss the requirement if the stack is the final tile in a row.
    Note, we do not have similar transitions from
    $\qhtile{\tile'}$
    since these states represent the relation between the next tile (stack) and
    the previous one, and thus need to be asserted at the end of a row too.

    The state
    $\qcn \in \sastates_1$
    above is an order-$1$ state from which we verify the column index is
    $\dimension$.
    To implement this state we need several intermediate states
    $\qcn^1, \ldots, \qcn^{2\cdot\numof} \in \sastates_1$.
    The first $\numof$ of these states allow any binary digit, while the final
    $\numof$, which read the bits encoding the column index, only allow $\bone$
    digits to occur.
    That is,
    \[
        \begin{array}{l}
            \qcn \satrancol{\spacer}{\emptyset} \set{\qcn^1},
            \\
            \qcn^1 \satrancol{\bzero}{\emptyset} \set{\qcn^2},
            \qcn^1 \satrancol{\bone}{\emptyset} \set{\qcn^2},
            \\
            \ldots,
            \\
            \qcn^\numof \satrancol{\bzero}{\emptyset} \set{\qcn^{\numof+1}},
            \qcn^\numof \satrancol{\bone}{\emptyset} \set{\qcn^{\numof+1}},
            \\
            \qcn^{\numof+1} \satrancol{\bone}{\emptyset} \set{\qcn^{\numof+2}},
            \ldots,
            \qcn^{2\cdot\numof-2}
                \satrancol{\bone}{\emptyset} \set{\qcn^{2\cdot\numof-1}},
            \qcn^{2\cdot\numof-1} \satrancol{\bone}{\emptyset} \emptyset
            \in
            \sadelta_1 \ .
        \end{array}
    \]

\item
    Finally we need to define transitions that check the vertical tiling
    relation.
    As before, we will begin with a transition reading the first spacer
    \[
        \qpropv \satran{\qany} \set{\qpropv'} \in \sadelta_2
    \]
    where
    $\qpropv, \qpropv' \in \sastates_2$.

    Next we will guess the pair
    $\tuple{\tile, \tile'} \in \vrel$
    that the next order-$1$ stack and its linked-to stack respectively will
    contain.
    For this we will need order-$2$ states
    $\qvtile{\tile}, \qvtilelink{\tile'} \in \sastates_2$
    for each
    $\tuple{\tile, \tile'} \in \vrel$
    from which we verify the next order-$1$ stack contain $\tile$ and the
    order-$1$ stack linked to from the $\spacer$ in this stack contains
    $\tile'$.
    The transitions we have are
    \[
        \qpropv'
            \satran{\qany} \set{\qpropv', \qvtile{\tile}, \qvtilelink{\tile'}}
        \in
        \sadelta_2
    \]
    where $\qpropv'$ appears on the right hand side to assert the vertical
    tiling relation over subsequent pairs of order-$1$ stacks.
    From each
    $\qvtile{\tile}$
    and
    $\qvtilelink{\tile'}$
    we have
    \[
        \qvtile{\tile} \satran{\qhastile{\tile}} \emptyset,
        \qvtilelink{\tile'} \satran{\qpass{\qvtilenext{\tile'}}} \emptyset,
        \qvtilenext{\tile'} \satran{\qhastile{\tile'}} \emptyset
        \in
        \sadelta_2
    \]
    where
    $\qvtilenext{\tile'} \in \sastates_2$
    are intermediate states for each required $\tile'$.
    Recall
    $\qpass{\qvtilenext{\tile'}}$
    will pass $\qvtilenext{\tile'}$ through the link on $\spacer$.

    Since the vertical tiling does not apply to the final tile in each column,
    we have transitions allowing the condition to be relaxed here.
    That is, we have, for each
    $\tuple{\tile, \tile'} \in \vrel$,
    \[
        \qpropv' \satran{\qrn} \emptyset,
        \qvtile{\tile} \satran{\qrn} \emptyset
        \in
        \sadelta_2 \ .
    \]
    Recall $\qrn$ verifies the row index is $\dimension$.
    Note, as soon as the automaton reaches the first order-$1$ stack containing
    a row index of $\dimension$, then all subsequent stacks will contain the
    same row index, hence the transition to $\emptyset$ allowing the condition
    to be disbanded.
    From
    $\qvtile{\tile}$
    we simply dismiss the requirement if the stack is the final tile in a column.
    Note, we do not have similar transitions from
    $\qvtilenext{\tile'}$
    since these states represent the relation between the next tile (stack) and
    the previous one via the collapse link, and thus need to be asserted at the
    end of a row too.
\end{enumerate}

%% file: conclusion.tex
\section{Conclusion}

We have shown that the problem of testing emptiness of stack automata is
NEXPTIME-complete for collapsible pushdown stacks.
In the case of \emph{annotated stacks} where stack characters are augmented
with further annotated stacks (instead of a link to a position elsewhere in the
stack), our proof does not carry through.
We conjecture that for annotated stacks the emptiness problem is
EXPTIME-complete but leave a proof of this for future work.